\documentclass[twocolumn,showpacs,showkeys,preprintnumbers]{revtex4}
\usepackage{amssymb}

\usepackage{amsfonts}

\usepackage{latexsym}

\usepackage{dcolumn}

\usepackage{amsmath}

\usepackage{graphicx}

\usepackage{psfrag,pstricks}

\newcommand{\eqn}{\ref}

\begin{document}

\title{Back-action ground state cooling of a micromechanical membrane via intensity-dependent interaction}

\author{ Sh. Barzanjeh$^{1}$ }
\email{shabirbarzanjeh@yahoo.com}

\author{M. H. Naderi$^{2}$}
\email{mhnaderi@phys.ui.ac.ir}
\author{ M. Soltanolkotabi$^{2}$}
\email{soltan@sci.ui.ac.ir}
\affiliation{ $^{1}$ Department of Physics, Faculty of Science, University of Isfahan, Hezar Jerib, 81746-73441, Isfahan, Iran\\
$^{2}$Quantum Optics Group, Department of Physics, Faculty of Science, University of Isfahan, Hezar Jerib, 81746-73441, Isfahan, Iran}

\date{\today}

\begin{abstract}

We propose a theoretical scheme to show the possibility of achieving the quantum ground state cooling of a vibrating micromechanical
membrane inside a high finesse optical cavity by back-action cooling approach. The scheme is based on an intensity-dependent
coupling of the membrane to the intracavity radiation pressure field. We find the exact expression for the position and momentum variances of the membrane by solving the linearized quantum Langevin equations in the steady-state, conditioned by the Routh-Hurwitz criterion. We show that by varying the Lamb-Dicke parameter and
the membrane's reflectivity one can effectively control the mean number of excitations of vibration of the membrane and also cool down the system to micro-Kelvin temperatures.
\end{abstract}

\maketitle

%TEXT--------------------------------------------------------------------------
\section{Introduction}\label{Introduction}
The coupling of mechanical motion of a mechanical resonator at the micro-and nano-meter scale to the electromagnetic degrees of freedom via radiation pressure\cite{cave,corbitt1,vitali1,kipp,fabre} is widely employed
for a large variety of applications\cite{cohadon}, more commonly as sensor to the detection of weak forces\cite{brad} and small displacements\cite{laha} or actuator in
integrated electrical, optical, and opto-electronical systems\cite{tang,brown}. Modification
of the resonator motion can be detected with high sensitivity by looking
at the radiation (or electric current) which interacted with the resonator. As an
example, by measuring the frequency shift induced
on the resonator one can detect the small masses. In principle, small displacements can be measured by detecting the corresponding phase shift of the light interacting with the resonator.

Moreover, the rapid progress of nano-technology has enabled fabrication of micro- or nanomechanical
resonators \cite{cle,blen} with high frequency, low dissipation and small mass. In this direction the most
experimental and theoretical efforts \cite{kle,gigan,bhatt,corbitt2,marshall,genes,yin} are
devoted to cooling and trapping such microresonators to their quantum ground state, but still in the
attempt to observe quantized mechanical motion, the thermal fluctuation has become a major
obstacle. The limited cooling efficiency and poor heat conduction at milli-Kelvin temperatures
of cryogenic refrigerators have stimulated a number of studies on the active cooling of
micro-mechanical resonators in both classical and quantum regimes\cite{cohadon,you}.
Very recently, various types of optomechanical system have demonstrated
significant cooling of the vibrational mode of a mechanical
resonator coupled to an optical cavity\cite{corbitt2,con,vit}. Among them, the so-called membrane-in-the middle geometry\cite{thompson,jay1,jay}in which the mechanical degree of freedom is a flexible, partially transparent dielectric membrane placed inside a Fabry-Perot cavity with fixed end mirrors has attracted much attention. This has the advantage of not having to combine the flexibility needed for the mechanical oscillator with the rigidity of a high-finesse cavity mirror. Although the membrane is nearly transparent, it couples to the optical cavity dispersively. This coupling is strong enough to laser-cool a 50-nm-thick dielectric membrane from room temperature(294 K) down to 7mK\cite{thompson}.

From experimental point of view the cooling of a membrane has been achieved by exploiting in
two different ways, back-action ground state cooling\cite{corbitt2,jay,bhatt,bra} and cold-damping quantum
feedback\cite{man,con,vit}. By cold-damping quantum
feedback, the oscillator position is measured
through a phase-sensitive detection of the cavity output and
the resulting photocurrent is used for a real-time correction
of the dynamics, and with the back-action cooling, the off-resonant operation
of the cavity results in a retarded back-action on the
mechanical system and hence in a self-modification of its
dynamics. In addition to this back-action effect,
the radiation pressure couples mechanical oscillator to the cavity mode. Thus in appropriate conditions the later, plays as an effective additional reservoir
for the oscillator. Therefore, strong radiation pressure coupling leads to significant cooling of the mechanical oscillator\cite{genes}.

Motivated by the above-mentioned studies on the cooling of mechanical oscillator, in the present paper
we deal with the study of cooling an optomechanical cavity with membrane-in-the middle geometry that consists of a high finesse cavity with two perfectly reflecting fixed end mirrors, and a partially reflective movable
middle mirror(such as a dielectric membrane). In this type of optomechanical structure, the radiation pressure and cavity detuning are periodic in the membrane displacement. This periodicity leads to an intensity-dependent interaction between the intracavity radiation pressure and the membrane's motion. We show that the presence of intensity-dependent interaction modifies the effective mechanical frequency and the effective mechanical damping of the membrane. It is also remarkable that in the steady-state condition, by controlling the system parameters, such as the membrane's reflectivity and the Lamb-Dicke parameter(LDP) the ground state cooling is approached. As we shall see by choosing proper values of these parameters the system can be cooled down to micro-Kelvin temperatures.

The paper is organized as follows. In Sec. II we derive an intensity-dependent
Hamiltonian describing the coupling of a micromechanical membrane to the radiation
pressure field through  \textit{j}-phonon excitations of the vibrational sideband. In Sec. III we derive the quantum Langevin equations(QLEs) of the system and linearize them around the semiclassical steady-state. In Sec. IV we investigate the effective frequency and the effective damping parameter of the mechanical oscillator. In Sec. V the back-action ground state cooling is discussed. Finally, a summary and some concluding remarks are
given in Sec. VI.

\section{Model Hamiltonian}

We consider a high finesse cavity as is shown in Fig.1,  which is detuned by the motion of a partially reflective membrane placed between two macroscopic, rigid, perfectly reflecting fixed end mirrors\cite{thompson,jay1,jay,borkje}. In this type of optomechanical system the coupling between the middle membrane and the optical cavity strongly depends on the position of the membrane. This position dependence results in a cavity detuning (for the fundamental mode
of motion of the membrane and of the cavity), which is a periodic function of the membrane displacement~$x$, i.e., $\omega_c(x)=(c/L)cos^{-1}[|r_c|cos(4\pi x/\lambda)]$ where $L$ and $r_c$ are the cavity length and the field reflectivity of the membrane, respectively, and also the position of membrane is calculated from antinode of cavity field. The Hamiltonian of the system is given by \cite{thompson,jay1}

\begin{eqnarray}\label{1}
H=\hbar \omega_c(x) a^{\dag}a +\hbar \omega_m  b^{\dag} b+i\hbar E(a^\dagger e^{-i\omega_lt}-a e^{i\omega_lt}),
\end{eqnarray}
where $ b$ and $ b^{\dagger}$($[ b, b^{\dag}]=1$) are the annihilation and creation operators for the membrane with oscillation frequency $\omega_m$ and $ a$ and $ a^{\dagger}$($[ a, a^{\dag}]=1$)are the annihilation and creation operators of the cavity mode with decay rate $\kappa$. The last term in the above Hamiltonian describes the input driving by a laser
with frequency $\omega_l$, where $E$ is related to the input laser
power $P_0$ by $|E|=\sqrt{2P_0\kappa/\hbar\omega_l}$. The fact that the motion of the membrane is quantized allows writing the operator of membrane's displacement $ x$ in terms of LDP, $\eta_0=\frac{4\pi}{\lambda}\sqrt{\frac{\hbar}{2m\omega_m}}$ ($\lambda$ and $m$ denote, respectively, the wavelength of the incident field and the motional mass of the membrane)as
\begin{eqnarray}\label{2}
 x=\eta_0(b^{\dag}+b).
\end{eqnarray}
 By expanding $\omega_c( x)$ in terms of $ b$ and $b^{\dagger}$ we obtain
\begin{widetext}
\begin{equation}\label{3}
\begin{array}{rcl}
\omega_c(x)=\frac{\pi c}{2L}-\frac{c}{2L}
\sum\limits_{m = 1}^{\infty} {\sum\limits_{k = 0}^{\frac{{m - 1}}{2}} {\frac{{\left| {r_c } \right|^m }}{m}} } \left( {\begin{array}{*{20}c}
   m  \\
   k  \\
\end{array}} \right)\frac{{(m - 1)!}}{{4^{m - 1} [(\frac{{m - 1}}{2})!]^2 }}\{ e^{i\eta_0 (m - 2k)( b+b^{\dagger})} + h.c.\},
\end{array}
\end{equation}
\end{widetext}
where $m=2l+1$ and $l$ is an integer number. By using the Baker-Campbell-Hausdorff theorem in Eq.(\eqn{3}) we may rewrite the Hamiltonian of Eq.(\eqn{1}) in the form
\begin{eqnarray}\label{4}
 H= H_0+ H^j_{int}+i \hbar E(a^\dagger e^{-i\omega_lt}-a e^{i\omega_lt}),
\end{eqnarray}
where
\begin{eqnarray}\label{5}
 H_0=\hbar \omega_0 a^{\dag}a +\hbar \omega_m  b^{\dag}b,
\end{eqnarray}\\
with $\omega_0=\pi c/(2L)$ as the natural frequency of the cavity without middle membrane, describes the free Hamiltonian of the quantized cavity field and the free motion of the mechanical degree of freedom, and
\begin{eqnarray}\label{6}
 H_{int}^j=\hbar a^{\dag} a[g^{*}_j f_j(n_b) b^{j}+g_j( b^{\dagger})^{j}f_j( n_b)]\nonumber,\,(j=0,1,2,..)\\
&&
\end{eqnarray}
where
\begin{equation}\label{9}
\begin{array}{rcl}
g_j=\frac{c}{2L}(i\frac{4\pi }{\lambda}\sqrt{\frac{\hbar}{2m\omega_m}})^j,
\end{array}
\end{equation}
and
\begin{widetext}
\begin{equation}\label{7}
\begin{array}{rcl}
f_j( n_b)=
\sum\limits_{m = 1} {\sum\limits_{k = 0}^{\frac{{m - 1}}{2}} {\frac{{\left| {r_c } \right|^m }}{m}} } \left( {\begin{array}{*{20}c}
   m  \\
   k  \\
\end{array}} \right)\frac{{(m - 1)!}}{{4^{m - 1} [(\frac{{m - 1}}{2})!]^2 }}\{ e^{-\frac{1}{2}(\eta_0)^{2} (m - 2k)^{2}}\}\frac{{n_b}!(m-2k)^{j}}{({n_b}+j)!}L^{j}_{{n_b}}[(\eta_0)^{2} (m - 2k)^{2}],
\end{array}
\end{equation}
\end{widetext}
with $ n_b= b^{\dagger} b$ and $L^{j}_{{n}}$ as the associated Laguerre polynomial, describes a nonlinear coupling of the radiation pressure field with the movable membrane through \textit{j}-phonon excitations of the vibrational sideband.

\section{Linearization of QLEs}

The dynamics of the system under consideration is governed by the fluctuation-dissipation
processes affecting both the optical and the
mechanical modes. To study these effects, let us consider the first excitation of the vibrational sideband by choosing $j=1$ in the Hamiltonian (\eqn{4}) with redefining the motional annihilation operator $b\rightarrow b e^{i \pi}$ and its conjugate $b^{\dagger}\rightarrow b^{\dagger} e^{-i \pi}$:
\begin{eqnarray}\label{8}
 H=\hbar \omega_0 a^{\dag} a +\hbar \omega_m  b^{\dag} b-\hbar \,g a^{\dag} a[
f( n_b) b+ b^{\dagger}f( n_b)]\nonumber\\
+i \hbar E(a^\dagger e^{-i\omega_lt}-a e^{i\omega_lt}),
\end{eqnarray}
where we have defined
\begin{eqnarray}
g\equiv g_1&=&i\frac{2\pi c}{L\lambda}\sqrt{\frac{\hbar}{2m\omega_m}},\nonumber\\
f( n_b)&\equiv &f_1( n_b).
\end{eqnarray}

The nonlinearity
function $f( n_b)$ determines the form of nonlinearity of the intensity-dependent of the coupling between the cavity field and the membrane. We point out that in the limit of very small values of LDP, $\eta_0$ and for certain values of the membrane reflectivity $r_c$ the nonlinearity function $f(n_b)$ reduces to unity. Fig.2(a) shows $f(n_b)$ as a function of $n_b$ for $8\eta_0$ and $r_c=0.99$, while Fig.2(b) displays the behavior of $f(n_b)$ versus $n_b$ for $10^{-4}\eta_0$ and $r_c=0.9$, where we have set $\eta_0=10^{-5}$. Obviously, for $f(n_b)=1$ the Hamiltonian (\eqn{8}) reduces to the Hamiltonian of the standard opto-mechanical system\cite{vitali1}. Therefore, the inherent nonlinearity of the model under consideration can be attributed to the parameters $\eta_0$ and $r_c$.

It should be noted that the experimental realization of the system under consideration shows $\eta_0<<1$,(e,g. for the experimental values given in Refs.\cite{thompson,jay1,jay}, we obtain $\eta_0\lesssim10^{-4}$). Therefore, one may keep terms up to first order in the phonons number $ n_b$ in Eq.(\eqn{7}) and approximate the nonlinearity function $f_{(j=1)}(n_b)=f(n_b)$ by expanding the associated Laguerre polynomial:
\begin{equation}\label{app}
\begin{array}{rcl}
f( n_b)\simeq \epsilon+\sigma n_b,
\end{array}
\end{equation}
where we have defined the following real parameters
\begin{widetext}
\begin{equation}
\begin{array}{rcl}
 \epsilon=\sum\limits_{m = 1} {\sum\limits_{k = 0}^{\frac{{m - 1}}{2}} {(m-2k)\frac{{\left| {r_c } \right|^m }}{m}} } \left( {\begin{array}{*{20}c}
   m  \\
   k  \\
\end{array}} \right)\frac{{(m - 1)!}}{{4^{m - 1} [(\frac{{m - 1}}{2})!]^2 }}\{ e^{-\frac{1}{2}(\eta_0)^{2} (m - 2k)^{2}}\},\\
\\
\sigma=\sum\limits_{m = 1} {\sum\limits_{k = 0}^{\frac{{m - 1}}{2}} {\frac{(i \eta_0)^2(m-2k)^3}{2!}\frac{{\left| {r_c } \right|^m }}{m}} } \left( {\begin{array}{*{20}c}
   m  \\
   k  \\
\end{array}} \right)\frac{{(m - 1)!}}{{4^{m - 1} [(\frac{{m - 1}}{2})!]^2 }}\{ e^{-\frac{1}{2}(\eta_0)^{2} (m - 2k)^{2}}\}.
\end{array}
\end{equation}
\end{widetext}

By substituting Eq.(\eqn{app}) into the Hamiltonian (\eqn{8}) we obtain,
\begin{equation}\label{ham4}
\begin{array}{rcl}
 H=H_{opt}-\hbar g\sigma a^{\dag} a(  n_b b+b^{\dagger} n_b)+i \hbar E(a^\dagger e^{-i\omega_lt}-a e^{i\omega_lt}),
\end{array}
\end{equation}
where $H_{opt}=\hbar \omega_0 a^{\dag} a +\hbar \omega_m  b^{\dag} b-\hbar \,g\epsilon a^{\dag} a(b+ b^{\dagger})$ denotes the Hamiltonian of standard optomechanical system and the second term is the contribution associated with the dependence of the cavity detuning on the position of the membrane. As we see from the Hamiltonian (\eqn{ham4}) the second term describes an intensity dependent coupling between the membrane's motion and the intracavity radiation pressure.

Now, let us consider the fluctuation-dissipation theorem for
the cavity field and the membrane's motion. For this purpose, We describe the effect of the fluctuations of the electromagnetic vacuum
and the Brownian noise associated with the coupling
of the oscillating mirror to its thermal environment within the input-output formalism of quantum
optics. For the given Hamiltonian (\eqn{ham4}) in the interaction picture this yields the following nonlinear QLEs
\begin{eqnarray}\label{lang1}
\dot{a}=-(i \Delta_0+\kappa)a+iga[\epsilon(b+b^{\dagger})&+&\sigma(n_b b+b^{\dagger}n_b)]\nonumber\\
&+&E+\sqrt{2\kappa} a_{in}(t),\nonumber\\
&&
\end{eqnarray}
\begin{equation}\label{lang2}
\begin{array}{rcl}
\dot{b}=-(i \omega_m+\gamma)b+iga^{\dagger}a[\epsilon+\sigma(2n_b+b^2)]+\sqrt{2\gamma} b_{in}(t),
\end{array}
\end{equation}
where $\Delta_0=\omega_c-\omega_0$ and $\gamma$ is the decay rate of the motional phonons of the membrane. The cavity field quantum vacuum fluctuation $a_{in}(t)$ satisfies the
Markovian correlation functions
\begin{eqnarray}\label{correfield}
\langle a_{in}(t)a_{in}^{\dagger}(t')\rangle &=&[\langle n_{th}\rangle+1]\delta(t-t'),\nonumber\\
\langle a_{in}^{\dagger}(t)a_{in}(t')\rangle &=&\langle n_{th}\rangle\delta(t-t'),\\
\langle a_{in}(t)a_{in}(t')\rangle&=&\langle a_{in}^{\dagger}(t)a_{in}^{\dagger}(t')\rangle=0,\nonumber
\end{eqnarray}
with the average thermal photon number $\langle n_{th}\rangle$. In our
study, we assume that the cavity is in zero temperature,
i.e., $\langle n_{th}\rangle = 0$. It means that the number of thermal photons
to be negligible at optical frequencies. Furthermore, the motional quantum fluctuation $b_{in}(t)$ satisfies the following relations
\begin{eqnarray}\label{correfield1}
\langle b_{in}(t)b_{in}^{\dagger}(t')\rangle&=&[\langle n_{b,th}\rangle+1]\delta(t-t'),\nonumber\\
\langle b_{in}^{\dagger}(t)b_{in}(t')\rangle&=&\langle n_{b,th}\rangle\delta(t-t'),\\
\langle b_{in}(t)b_{in}(t')\rangle&=&\langle b_{in}^{\dagger}(t)b_{in}^{\dagger}(t')\rangle=0,\nonumber
\end{eqnarray}
where $\langle n_{b,th}\rangle$ is the mean number of phonons in the absence of optomechanical coupling, determined by the temperature of the mechanical bath $T$
\begin{eqnarray}
\langle n_{b,th}\rangle=\frac{1}{e^{\frac{\hbar \omega_m}{k_B T}}-1}.
\end{eqnarray}

Analyzing the quantum dynamics of the full nonlinear system is difficult, so we linearize the QLEs around the semiclassical fixed points. That is, we decompose each operator in Eqs.(\eqn{lang1}) and (\eqn{lang2}) as the
sum of its steady-state value and a small fluctuation, e.g., $b=b_s+\delta b$ and $a=a_s+\delta a$.  This decouples our system into a set of nonlinear algebraic equations for the steady-state values and a set of QLEs for the fluctuation operators. We point out that the linearization of the QLEs is compatible with our aim i.e., the ground state cooling because as is shown in Ref.\cite{vitali1}, significant  cooling is approached when the radiation pressure coupling is strong, which needs very intense intracavity fields i.e., $|\alpha_s|>>1$.  Thus the steady-state equations corresponding to the Eqs.(\eqn{lang1}) and (\eqn{lang2}) are given by
\begin{equation}\label{ss1}
\begin{array}{rcl}
a_s=\frac{E}{\sqrt{\kappa^2+\Delta^2}},\,\,b_s=\frac{G a_s}{\omega_m-i\gamma},
\end{array}
\end{equation}
where we have defined the effective detuning $\Delta$ and the effective coupling constant $G$ as
\begin{eqnarray}\label{ss1}
\Delta&=&\Delta_0-2g Re(b_s)(\epsilon+\sigma|b_s|^2),\\\nonumber
G&=&2a_s g[\epsilon+\sigma( b_s^2+2|b_s|^2)].
\end{eqnarray}
By eliminating
the steady-state contribution and linearizing the resulting
equations for the fluctuations, we obtain the exact QLEs
for the fluctuations of the quadrature operators
\begin{equation}\label{lanf1}
\begin{array}{rcl}
\delta \dot{X}&=&\Delta \delta Y-\kappa \delta X+\sqrt{2\kappa}X_{in},
\end{array}
\end{equation}
\begin{equation}\label{lanf2}
\begin{array}{rcl}
\delta \dot{Y}&=&-\Delta \delta X-\kappa \delta Y+(G_R\delta q+G_I\delta p)+\sqrt{2\kappa}X_{in},
\end{array}
\end{equation}
\begin{equation}\label{lanf3}
\begin{array}{rcl}
\delta \dot{q}&=&\Omega_1 \delta p-\Gamma_1\delta q-G_I\delta X+\sqrt{2\gamma}q_{in},
\end{array}
\end{equation}
\begin{equation}\label{lanf4}
\begin{array}{rcl}
\delta \dot{p}&=&-\Omega_2 \delta q-\Gamma_2\delta p+G_R \delta x+\sqrt{2\gamma}p_{in},
\end{array}
\end{equation}
where $G_R$ and $G_I$ are, respectively, the real part and the imaginary part of $G$ and we have symmetrized the fluctuation operators as
\begin{eqnarray}\label{eqmotion1}
\delta X=\frac{1}{\sqrt{2}}(\delta a+\delta a^{\dagger}),
\delta Y=\frac{1}{\sqrt{2i}}(\delta a-\delta a^{\dagger}),\\
\delta q=\frac{1}{\sqrt{2}}(\delta b+\delta b^{\dagger}),
\delta p=\frac{1}{\sqrt{2i}}(\delta b-\delta b^{\dagger}).
\end{eqnarray}
Furthermore we have defined
\begin{eqnarray}\label{para}
\Omega_m&=&\omega_m-4g\sigma a_s^2 Re(b_s),\nonumber\\
\Omega_1&=&\Omega_m+2g\sigma a_s^2 Re(b_s),\\
\Omega_2&=&\Omega_m-2g\sigma a_s^2 Re(b_s),\nonumber\\
\Gamma_1&=&\gamma+2g\sigma a_s^2 Im(b_s),\nonumber\\
\Gamma_2&=&\gamma-2g\sigma a_s^2 Im(b_s).\nonumber
\end{eqnarray}
We point out that in the limit of $\sigma\longrightarrow0$, Eqs.(\eqn{lanf1})-(\eqn{lanf4}) reduce to the QLEs of standard optomechanical system \cite{genes}. These equations show that the nonlinear term in the Hamiltonian (\eqn{ham4}) leads to additional parts in QLEs. In Eq.(\eqn{lanf2}), the term $G_I\delta p$ and in Eq.(\eqn{lanf3}), the term $-G_I\delta X$ directly depend upon the parameter $\sigma$, which introduces further coupling in QLEs and this shows that the nonlinearity is responsible for the appearance of the imaginary part of the effective coupling constant $G$. It is evident from Eqs.(\eqn{para}) that the modification due to the existence of nonlinearity is not limited just for $G$; the detuning and the damping parameters in Eq.(\eqn{para}) are also affected by the parameter $\sigma$.

\section{Effective frequency and effective damping parameter of the membrane}
In this section we evaluate the effective frequency $\omega_{eff}$ and
the effective damping  rate $\gamma_{eff}$ of the membrane in the system under consideration. For this purpose, we solve the linearized QLEs
for the fluctuations in the displacement operator as

\begin{equation}\label{qfun}
\begin{array}{rcl}
\delta q(\omega)=\chi(\omega)\textit{f}_T(\omega),
\end{array}
\end{equation}
where
\begin{equation}\label{qfun1}
\begin{array}{rcl}
\textit{f}_T(\omega)=\frac{1}{2\pi}\int_{-\infty}^{\infty}dt e^{i\omega t}\textit{f}_T(t),
\end{array}
\end{equation}
is the Fourier transformation of the fluctuations in the total force acting on the
membrane, which includes a radiation vacuum and a Brownian motion
component. Here $\chi(\omega)$ describes the mechanical susceptibility of the membrane, given by
\begin{equation}\label{lanf14}
\begin{array}{rcl}
\chi^{-1}(\omega)=\frac{1}{\Omega_1}[\Omega_1 \Omega_2+(\Gamma_1-i\omega)(\Gamma_2+i\omega)-\Omega_1I(\omega)],
\end{array}
\end{equation}
with
\begin{equation}\label{lanf14}
\begin{array}{rcl}
I(\omega)=\frac{\Delta[G_R+G_I(\Gamma_1-i\omega)/\Omega_1][G_R-G_I(\Gamma_2-i\omega)/\Omega_1]}{(k-i\omega)^2+\Delta^2-\Delta G_I^2/\Omega_1}.
\end{array}
\end{equation}
The mechanical susceptibility of the membrane can be considered as the susceptibility
of an oscillator with effective resonance frequency and
effective damping rate, respectively, given by

\begin{equation}\label{fre}
\begin{array}{rcl}
\omega_{eff}^2(\omega)=\Gamma_1\Gamma_2+\Omega_1\Omega_2-\frac{\Delta[\mu_1+\mu_2\omega^2-G_I^2\omega^4]}{\Omega_1[\kappa^2+(\omega-\Delta')^2]
[\kappa^2+(\omega+\Delta')^2]},
\end{array}
\end{equation}
\begin{equation}\label{damp}
\begin{array}{rcl}
\frac{\gamma_{eff}(\omega)}{2}&=&\gamma+\frac{\Delta\big[\mu_3-(\kappa+\gamma)G_I^2\omega^2\big]}{\Omega_1[\kappa^2+(\omega-\Delta')^2][\kappa^2+(\omega+\Delta')^2]},
\end{array}
\end{equation}
where $\Delta'^2=\Delta^2-\Delta G_I^2/\Omega_1$ and we have defined
\begin{eqnarray}
\mu_1&=&(\kappa^2+\Delta'^2)(\Gamma_1 G_I+\Omega_1 G_R)(\Omega_1 G_R-\Gamma_2 G_I),\nonumber\\
\mu_2&=&(\kappa^2 + \Gamma_1 \Gamma_2 -2 \kappa (\Gamma_1 + \Gamma_2) + \Delta'^2)
G_I^2\nonumber\\
 &+& (\Gamma_2 - \Gamma_1) \Omega_1 G_I G_R- \Omega_1^2 G_R^2,\\
\mu_3&=&\gamma G_I^2(\kappa^2+\Delta'^2)+\nonumber\\
&&\kappa\Big[\Omega_1^2 G_R^2+\Omega_1G_I G_R(\Gamma_1-\Gamma_2)- G_I^2\Gamma_1\Gamma_2\Big].\nonumber
\end{eqnarray}
The modification of the mechanical frequency due to the radiation pressure shown by Eq.(\eqn{fre}) is the so-called \textit{optical spring effect}\cite{genes}. The existence of nonlinearity in the Hamiltonian (\eqn{ham4}) leads to the appearance of higher order of
system response frequency $\omega$, in contrast to the results of the
linear coupling theory\cite{genes}. It is evident that in the limit of $\sigma\longrightarrow0$ the terms $G_I^2\omega^4/\Omega_1^2$ in Eq.(\eqn{fre}) and $(\kappa+\gamma)(G_I/\Omega_1)^2\omega^2$ in Eq.(\eqn{damp}) are removed. Under this condition the effective frequency and the effective damping parameter reduce to the corresponding parameters in standard optomechanical system(See Eqs.(18) and (19) in Ref.\cite{genes}).
Fig.3(a) shows the normalized effective frequency as a function of the normalized system response frequency $\omega/\omega_m$ at $\Delta=\omega_m$ for different values of LDP. As is seen, LDP alters significantly the effective frequency. The normalized effective damping rate versus the normalized system response frequency $\omega/\omega_m$ at $\Delta=\omega_m$ have been plotted in Fig.3(b). As is seen, by increasing LDP the effective damping rate increases. The significate increasing of the effective mechanical damping rate is the
basis of the cooling process. The ground state cooling can be obtained only if the initial mean thermal
excitation number $\bar{n}=[exp(\hbar \omega_m/k_B T)-1]^{-1}$ is not prohibitively
large. This can be approached for large values of the effective mechanical damping rate. As is clear from Fig.3(b) for positive $\Delta$ and by choosing proper values of LDP the effective mechanical damping rate is
significantly increased.

\section{Back-action ground state cooling in the presence of intensity-dependent coupling}
In this section we study the effective ground state cooling of the membrane's motion. We first evaluate the mean energy in
the steady-state by defining the system's potential as
 \begin{equation}\label{enr}
\begin{array}{rcl}
U=\frac{\hbar \omega_m}{2}\Big(\langle\delta q(t)^2\rangle+\langle\delta p(t)^2\rangle\Big)=\hbar \omega_m(n_{eff}+\frac{1}{2}).
\end{array}
\end{equation}\\
The ground state is approached for $n_{eff}\simeq0$ or $U=\hbar \omega_m/2$. This is obtained if $\langle\delta q(t)^2\rangle\simeq\langle\delta p(t)^2\rangle\simeq1/2$ occurring in steady-state. The system reaches a steady-state only if it is stable. In this situation all the poles of the effective susceptibility $\chi$
lie in the lower complex half-plane. The stability conditions can be attained by applying the Routh-Hurwitz criterion \cite{rh} which gives the following stability expressions
\begin{eqnarray}\label{ste}
M_1&=&(\kappa^2+\Delta^2)(\Omega_1\Omega_2+\Gamma_1\Gamma_2)\\
&&-\Delta[\Omega_2G_I^2+\Omega_1G_R^2+G_RG_I(\Gamma_1-\Gamma_2)]>0,\nonumber\\
\nonumber\\
M_2&=&\Delta^4+\Delta^2[\gamma^2-2\Gamma_1\Gamma_2+4\kappa\gamma+2(\kappa^2-\Omega_1\Omega_2)]\nonumber\\
&&+\Delta\frac{2(\kappa+\gamma)^2}{\kappa\gamma}[\Omega_2G_I^2+\Omega_1G_R^2+G_RG_I(\Gamma_1-\Gamma_2)]\nonumber\\
&&+[\Omega_1\Omega_2+(\Gamma_1+\kappa)(\Gamma_2+\kappa)]^2>0.\nonumber
\end{eqnarray}

Now, we turn our attention to Eq.(\eqn{qfun}) in which $\textit{f}_T(\omega)$ is the fluctuations in the total force acting on the membrane's motion and it has the following form
\begin{widetext}
\begin{eqnarray}\label{no1}
\textit{f}_T(\omega)=\sqrt{2\gamma} p_{in}&+&\sqrt{2\gamma} \frac{q_{in}}{\Omega_1}\Big[\Gamma_2-i \omega-\frac{\Delta[\Omega_1G_R-G_I(\Gamma_2-i\omega)]}{\Omega_1(\kappa-i\omega)^2+\Omega_1\Delta^2-\Delta G_I^2}\Big]\nonumber\\
&&+\sqrt{2\kappa}\Big[\frac{\Omega_1G_R-G_I(\Gamma_2-i\omega)}{\Omega_1(\kappa-i\omega)^2+\Omega_1\Delta^2-\Delta G_I^2}\Big]\Big[\Delta Y_{in}+(\kappa-i\omega)X_{in}\Big],
\end{eqnarray}
\end{widetext}
where the first and second terms show the fluctuations in the membrane's motion, and the third term describes the fluctuations in the intracavity field. By using Eq.(\eqn{qfun}) we obtain
\begin{eqnarray}\label{qf}
\langle\delta q(\omega)\delta q(\omega')\rangle= \chi(\omega)\chi(\omega')\langle\textit{f}_T(\omega)\textit{f}_T(\omega')\rangle,
\end{eqnarray}
and by inverse Fourier transforming Eq.(\eqn{qf}) we have
\begin{eqnarray}\label{noiseope}
\langle\delta q(t)\delta q(t')\rangle=\frac{ 1}{2\pi}&&\int_{-\infty}^{\infty}d\omega \int_{-\infty}^{\infty} d\omega' \times\\\nonumber
 &&e^{-i\omega t}e^{i\omega' t'}\chi(\omega)\chi(\omega')\langle\textit{f}_T(\omega)\textit{f}_T(\omega')\rangle.
\end{eqnarray}
By using Eq.(\eqn{no1}) and the quantum fluctuation relations (\eqn{correfield}) and (\eqn{correfield1}) we find
\begin{widetext}\label{noise3}
\begin{eqnarray}
\langle\textit{f}_T(\omega)\textit{f}_T(\omega')\rangle&=&\Big[\gamma(2n_{b,th}+1)+\frac{\gamma(2n_{b,th}+1)}{\Omega_1}\Big|\Gamma_2-i\omega-
\frac{\Delta C(\omega)}{D(\omega)}\Big|^2\nonumber\\
&+&2\gamma\Big(\omega+\Delta Im[\frac{C(\omega)}{D(\omega)}]\Big)+\kappa\Big|\frac{C(\omega)}{D(\omega)}\Big|^2(\Delta^2+\kappa^2+\omega^2+\Delta \omega)\Big]\delta(\omega+\omega'),
\end{eqnarray}
\end{widetext}
where we have defined
\begin{eqnarray}\label{qf2}
C(\omega)&&=G_R-G_I(\Gamma_2-i\omega)/\Omega_1,\nonumber\\
D(\omega)&&=(\kappa-i\omega)^2+\Delta'^2.
\end{eqnarray}
The integral of Eq.(\eqn{noiseope}) for the displacement variance can be solved exactly when the stability conditions given by Eqs.(\eqn{ste}) are satisfied. Performing the integral and using $\chi(-\omega)=\chi^*(\omega)$ with setting $t=t'$, one gets the final expression for
the time dependent position variance
\begin{eqnarray}\label{q22}
\langle\delta q^2\rangle=\gamma(2n_{b,th}+1)N_1+\kappa N_2,
\end{eqnarray}
where
\begin{eqnarray}\label{qf2}
2N_1&=&\Omega_1^2\Big(\frac{a_1b_2-a_3b_1}{M_2}
+\frac{b_3(a_3-a_1a_2)}{M_2M_1}\Big)\nonumber\\
&+&\frac{a_1d_2-a_3d_1+a_2a_3-a_1a_4}{M_2}
+\frac{d_3(a_3-a_1a_2)}{M_2M_1},\nonumber\\
2N_2&=&\Omega_1^2\Big(\frac{a_1c_2-a_3c_1}{M_2}+\frac{c_3(a_3-a_1a_2)}{M_2M_1}\Big).
\end{eqnarray}
By applying the same procedure, we obtain the following expression for the momentum variance
\begin{eqnarray}\label{p22}
\langle\delta p^2\rangle=\gamma(2n_{b,th}+1)N'_1+\kappa N'_2,
\end{eqnarray}
where
\begin{eqnarray}\label{qf2}
2N'_1&=&\Omega_2^2\Big(\frac{a_1b'_2-a_3b_1}{M_2}
+\frac{b'_3(a_3-a_1a_2)}{M_2M_1}\Big)\nonumber\\
&+&\frac{a_1d'_2-a_3d_1+a_2a_3-a_1a_4}{M_2}
+\frac{d'_3(a_3-a_1a_2)}{M_2M_1},\nonumber\\
2N'_2&=&\Omega_2^2\Big(\frac{a_1c'_2-a_3c'_1}{M_2}+\frac{c'_3(a_3-a_1a_2)}{M_2M_1}\Big).
\end{eqnarray}
The explicit expressions for the parameters $a_i, b_i, c_i,d_i$ and $b'_i, c'_i,d'_i $ are given in the appendix.

The effective phonon number of the membrane's motion can be defined from Eq.(\eqn{enr})
\begin{equation}\label{noise2}
\begin{array}{rcl}
n_{eff}=\frac{1}{2}(\langle\delta q(t)^2\rangle+\langle\delta p(t)^2\rangle-1).
\end{array}
\end{equation}
As we mentioned, the ground state cooling is approached if $n_{eff}<1$ which is not the only condition for cooling. In order
to get the quantum ground state cooling it is also necessary that the uncertainly principle is stratified for both momentum and displacement variances. This means that both variances have
to tend to $\langle\delta q^2\rangle\simeq\langle\delta p^2\rangle\simeq1/2$, therefore energy equipartition has to be
satisfied in the optimal regime close to the ground state. We point out that, in general, one does not
have energy equipartition because these variances are not equal $\langle\delta q(t)^2\rangle\neq\langle\delta p(t)^2\rangle$.

In order to have an intuitive picture, we have plotted the both variances in Fig.4 for different values of LDP and for the experimentally feasible
parameters (see Table.1 in Ref.\cite{thompson} and also \cite{jay1} ), i.e., a cantilever with \textit{motional mass} $m=50pg$$, \omega_m/2\pi=10^5Hz$, $r_c=0.999$  and $Q=1.2\times 10^7$ placed inside an optical cavity with length $L=6.7cm$ and damping rate $\kappa=0.047\omega_m$ driven by a laser with wavelength $\lambda_{0}=1064nm$ and power $P_c=50\mu W$. It is evident from Fig. 4 that by varying LDP one can access the ground state cooling condition $\langle \delta q^2\rangle\simeq\langle\delta p^2\rangle\simeq1/2$. As is seen from fig.4(c)for $\eta_0$ around $\Delta=1.1\omega_m$ one has $\langle\delta q^2\rangle\simeq\langle\delta p^2\rangle\simeq1/2$. This means that LDP may significantly alter the cooling process and it can be used as an additional control parameter to optimize the best condition for the ground state cooling in the system.

Fig.5(a) shows the effective phonon number as a function of detuning for three different values of LDP. It is clear that with the increasing value of LDP the minimum value of $n_{eff}$ is shifted toward zero, such that for $\eta_0$ we have $n_{eff}\simeq0$.
To investigate the effect of the membrane's reflectivity on ground state cooling we have plotted the effective phonon number as a function of detuning in Fig.5(b). It is interesting that increasing of $r_c$ significantly decreases the effective phonon number. One can also define an effective
temperature with respect to the effective mean excitation
number $n_{eff}$ as
\begin{equation}\label{temp}
\begin{array}{rcl}
T_{eff}=\frac{\hbar \omega_m}{k_B\, ln(1+1/n_{eff})}.
\end{array}
\end{equation}
we have plotted the effective temperature for experimental values given in Fig. 4 and for the initial reservoir temperature $T=0.4K$ corresponding to $\bar n\simeq83306$ against the normalized detuning and for different values of LDP. It is remarkable that the minimum achievable temperature is $T_{eff}\simeq1\mu K$. Therefore by controlling LDP and $r_c$ one can control the minimum value of the effective temperature of the system.
\section{summary and conclusions}
In this paper we have introduced a physical scheme to develop the ground state cooling of a mechanical membrane. As we have seen, the dependence of cavity detuning to the membrane displacement leads to an intensity-dependent interaction between the intracavity radiation pressure and the membrane's motion. Consequently, an additional nonlinear term is appeared in the Hamiltonian of standard optomechanical system. Existence of this part in the Hamiltonian alters the dynamical behaviour of the system by introducing extra coupling terms in QLEs. We have derived the effective mechanical susceptibility and have studied the effect of LDP on the effective mechanical frequency and the effective damping parameter.\\
As we have also seen, the additional nonlinear term enhances significantly the back-action ground state cooling process to obtain smaller effective temperature. By varying LDP and membrane's reflectivity one can optimize the cooling process to achieve the ground state. We have found that for initial reservoir temperature $T=0.4K$(corresponding to $\bar n\simeq83306$) and in the experimentally accessible
parameter regimes, one can observe the mean mechanical excitation number close to zero with corresponding effective temperature around micro-Kelvin.
\section*{Acknowledgments}
The authors would like to express their gratitude to the
referee, whose valuable comments have improved the
paper. They are also grateful to the Office of Graduate
Studies of the University of Isfahan for their support.
\appendix*
\section{solution of rational integral of Eq.(\eqn{noiseope})}
when the stability condition is satisfied, according the Routh-Hurwitz criterion one can solve exactly the integral of rational function of Eq.(\eqn{noiseope})to find the position and momentum variances of the membrane given, respectively, by Eq.(\eqn{q22}) and by Eq.(\eqn{p22}) in which
\begin{eqnarray}\label{qf2}
a_1&=&2(\gamma+\kappa),\,a_2=-(\Delta^2+\kappa^2+\Gamma_1\Gamma_2+\Omega_1\Omega_2+4\gamma\kappa),\nonumber\\
a_3&=&-2[\gamma(\Delta^2+\kappa^2)+\kappa(\Gamma_1\Gamma_2+\Omega_1\Omega_2)],\nonumber\\
b_1&=&1,\,b_2=2(\kappa^2-\Delta_3^2),\,\,b_3=(\kappa^2+\Delta_3^2)^2,\nonumber\\
c_1&=&(\frac{G_I}{\Omega_1})^2,\,c_2=(\Delta^2+\kappa^2)c_1+(G_R-\frac{G_I\Gamma_2}{\Omega_1})^2,\nonumber\\
c_3&=&(\Delta^2+\kappa^2)(G_R-\frac{G_I\Gamma_2}{\Omega_1})^2,\\
d_1&=&2(\kappa^2-\Delta^2)+\Gamma_2^2,\nonumber\\
d_2&=&(kc^2+\Delta^2)^2 +2 \Gamma_2^2( kc^2 -\Delta^2)+
 2 (2 kc + \Gamma_2) \Delta G_I G_R,\nonumber\\
d_3&=&\Big[(\kappa^2+\Delta^2)\Gamma_2-\Delta G_I G_R\Big]^2,\nonumber
\end{eqnarray}
and
\begin{eqnarray}\label{qf2}
b'_2&=&2(\kappa^2-\Delta_3'^2),\,\,b'_3=(\kappa^2+\Delta_3'^2)^2,\nonumber\\
c'_1&=&(\frac{G_R}{\Omega_2})^2,\,c'_2=(\Delta^2+\kappa^2)c'_1+(G_I+\frac{G_R\Gamma_1}{\Omega_2})^2,\nonumber\\
c'_3&=&(\Delta^2+\kappa^2)(G_I+\frac{G_R\Gamma_1}{\Omega_2})^2,\\
d'_2&=&(kc^2+\Delta^2)^2 +2 \Gamma_1^2( kc^2 -\Delta^2)-
 2 (2 kc + \Gamma_1) \Delta G_I G_R,\nonumber\\
d'_3&=&\Big[(\kappa^2+\Delta^2)\Gamma_1+\Delta G_I G_R\Big]^2,\nonumber
\end{eqnarray}
with $\Delta_3'^2=\Delta^2-\frac{\Delta G_R^2}{\Omega_2}$.
%Ref-----------------------------------------------------------------------
%\section*{References}
\bibliographystyle{apsrev4-1}

\begin{thebibliography}{10}

\expandafter\ifx\csname natexlab\endcsname\relax\def\natexlab#1{#1}\fi
\expandafter\ifx\csname bibnamefont\endcsname\relax
  \def\bibnamefont#1{#1}\fi
\expandafter\ifx\csname bibfnamefont\endcsname\relax
  \def\bibfnamefont#1{#1}\fi
\expandafter\ifx\csname citenamefont\endcsname\relax
  \def\citenamefont#1{#1}\fi
\expandafter\ifx\csname url\endcsname\relax
  \def\url#1{\texttt{#1}}\fi
\expandafter\ifx\csname urlprefix\endcsname\relax\def\urlprefix{URL }\fi


\bibitem {cave} C. M. Caves, \textit{Phys. Rev. Lett.} \textbf{45}, 75(1980).
\bibitem {corbitt1} T. Corbitt \textit{et al.}, \textit{Phys. Rev. A} \textbf{74}, 021802(2006).

\bibitem{vitali1}D. Vitali, S. Gigan, A. Ferreira, H. R. Böhm, P. Tombesi, A.
Guerreiro, V. Vedral, A. Zeilinger, and M. Aspelmeyer, \textit{Phys.
Rev. Lett.} \textbf{98}, 030405 (2007);
X. Zou and W. Mathis, \textit{Phys. Lett. A} \textbf{324}, 484 (2004);
 A. N. Cleland and M. R. Geller, \textit{Phys. Rev. Lett.} \textbf{93}, 070501
(2004).

\bibitem {kipp} T.J. Kippenberg and K. J. Vahala, \textit{Opt.Express} \textbf{15}, 17172(2007).
\bibitem {fabre} C. Fabre, M. Pinard, S. Bourzeix, A.
Heidmann, E. Giacobino, and S. Reynaud, \textit{Phys. Rev. A} \textbf{49}, 1337
(1994);
K. Jacobs, P. Tombesi, M. J. Collett, and D. F. Walls, \textit{Phys.
Rev. A }\textbf{49}, 1961 (1994).
\bibitem {cohadon} P. F. Cohadon, A. Heidmann, and M. Pinard, \textit{Phys. Rev. Lett.} \textbf{83}, 3174(1999);
Y. Hadjar \textit{et al.}, \textit{Europhys. Lett.} \textbf{47}, 545(1999);
S. Mancini, V. Giovannetti, D. Vitali, and P. Tombesi, \textit{Phys. Rev. Lett.} \textbf{88}, 120401(2002);
C. H. Metzger and K. Karrai, \textit{Nature (London)} \textbf{432}, 1002
(2004);
B. S. Sheard \textit{et al.}, \textit{Phys. Rev. A} \textbf{69}, 051801(R)(2004);
H. Rokhsari, T.J. Kippenberg, T. Carmon, and K. J. Vahala, \textit{IEEE J. Quantum Electron.} \textbf{12}, 96(2006);
O. Arcizet \textit{et al.}, \textit{Nature (London)} \textbf{444}, 71 (2006);
A. Schliesser \textit{et al.}, \textit{Phys. Rev. Lett.} \textbf{97}, 243905 (2006);
A. DiVirgilio \textit{et al.}, \textit{Phys. Rev. A} \textbf{74}, 013813(2006);
M. Bhattacharya and P. Meystre, \textit{Phys. Rev. Lett}. \textbf{99},
073601 (2007);
M. Ludwig, B. Kubala and F. Marquardt, \textit{New. J. Phys.} \textbf{10}, 095013(2008);
A. Xuereb \textit{et al.}, \textit{Phys. Rev. A} \textbf{79}, 053810(2009).
\bibitem {brad} C. Bradaschia \textit{et al.}, \textit{Nucl. Instrum. Methods Phys. Res. A}
\textbf{289}, 518 (1990); A. Abramovici \textit{et al.,} \textit{Science} \textbf{256}, 325(1992);
 P. Fritschel, \textit{Proc. SPIE }\textbf{4856}, 282 (2003).
\bibitem {laha}M. D. LaHaye, O. Buu, B. Camarota, and K. C. Schwab, \textit{Science}
\textbf{304}, 74 (2004).
\bibitem {tang}M. Tsang, \textit{Phys. Rev. A} \textbf{81}, 063837 (2010).
\bibitem {brown}K. R. Brown  \textit{et al.},  \textit{Phys. Rev. Lett.} \textbf{99}, 137205 (2007).
\bibitem {cle}A. N. Cleland \textit{Foundations of Nanomechanics: From Solid-state Theory to Device Applications }(
Springer-Verlag, 2002).
\bibitem{blen}M. Blencowe \textit{Phys. Rep.} \textbf{395} 159(2004 ).

\bibitem {kle}D. Kleckner and D. Bouwmeester, \textit{Nature (London)} \textbf{444}, 75 (2006).
\bibitem {gigan}S. Gigan \textit{et al.}, \textit{Nature (London)}  \textbf{444}, 67 (2006).
\bibitem {bhatt}M. Bhattacharya, H. Uys, and P. Meystre, \textit{Phys. Rev. A}
\textbf{77}, 033819 (2008).

\bibitem {marshall} W. Marshall, C. Simon, R. Penrose, and D. Bouwmeester, \textit{Phys. Rev. Lett.} \textbf{91}, 130401 (2003).
\bibitem {corbitt2}T. Corbitt \textit{et al.}, \textit{Phys. Rev. Lett.} \textbf{98}, 150802 (2007).
\bibitem {genes} C. Genes \textit{et al.,} \textit{Phys. Rev. A} \textbf{77},  033804(2008).
\bibitem {yin}Y-D. Wang, K. Semba and H. Yamaguchi,  \textit{New J. Phys.} \textbf{10}, 043015(2008).
\bibitem {you} J. Q. You \textit{et al.,}\textit{Phys. Rev. Lett.} \textbf{100}, 047001(2008).
\bibitem {con}J.-M. Courty, A. Heidmann, and M. Pinard, \textit{Eur. Phys. J. D} \textbf{17},
399 (2001).
\bibitem {vit}D. Vitali, S. Mancini, L. Ribichini, and P. Tombesi, \textit{Phys. Rev.
A} \textbf{65}, 063803 (2002).
\bibitem {thompson}J. D. Thompson \textit{et al.}, \textit{Nature (London)} \textbf{452}, 72
(2008).
\bibitem{jay1}A. M. Jayich \textit{et al.}, \textit{New. J. Phys.} \textbf{10}, 095008 (2008).
\bibitem {jay}
M. Bhattacharya and P. Meystre, e-print arXiv: 0803.1219v1;
J. C. Sankey, C. Yang, B. M. Zwickl, A. M. Jayich, and J. G. Harris, e-print arXiv:1002.4158.
\bibitem {bra}V. B. Braginsky, S. E. Strigin, and S. P. Vyatchanin, \textit{Phys. Lett.
A} \textbf{287}, 331 (2001).
\bibitem {man}S. Mancini, D. Vitali, and P. Tombesi, \textit{Phys. Rev. Lett.} \textbf{80},688 (1998);

\bibitem {borkje} K. Borkje \textit{et al}.,e-print arXiv:1004.3587.


\bibitem {rh}I. S. Gradshteyn and I. M. Ryzhik, \textit{Table of Integrals, Series
and Products} (Academic Press, Orlando, 1980); A. Hurwitz, \textit{Selected Papers on Mathematical Trends in Control Theory}, edited by R. Bellman and R. Kabala (Dover, New York, 1964).
\end{thebibliography}

\vskip 36cm
\section*{Figure Captions}
\textbf{Fig.1:} A high-finesse optical cavity with two rigid end mirrors and a dielectric membrane centered at the antinode of cavity field.\\
\\
\textbf{Fig.2:} The nonlinearity function $f(n_b)$ as a function of phonon number $n_b$ for: (a) $r_c=0.99$, $8\eta_0$. (b)  $r_c=0.9$, $10^{-4}\eta_0$. Here we have set $\eta_0=10^{-5}$.\\
\\
\textbf{Fig.3:}(a)The effective mechanical frequency of Eq.(\eqn{fre}) and
(b) The effective mechanical damping rate of Eq.(\eqn{damp}) versus
normalized frequency. Parameter values are\cite{thompson,jay1} $\omega_m/2\pi=10^5Hz, L=7cm, P_0=61\mu W, \kappa=0.051 \omega_m, Q=1.2\times10^7$ ,motional mass $m=0.5pg$ and $\eta_0=6.8\times10^{-7}$(\textit{i})$0.65\eta_0$(red line),(\textit{ii})$0.7\eta_0$(green line) and (\textit{iii})$\eta_0$(blue line).\\
\\
\textbf{Fig.4:} the membrane's position variance $\langle\delta q^2\rangle$(red line) and the membrane's momentum variance $\langle\delta p^2\rangle$(blue line) versus the
normalized effective detuning for the initial reservoir temperature $T=0.4$ corresponding to $n_{eff}\simeq83306$ and for different values of LDP:(a)$0.85\eta_0$ (b)$0.9\eta_0$ and (c)$\eta_0$.\\
\\
\textbf{Fig.5:} The effective mean excitation $n_{eff}$ versus
the normalized effective detuning for the initial reservoir temperature $T=0.4$ corresponding to $n_{eff}\simeq83306$ (a)for different values of LDP: $0.85\eta_0$(red line), $0.9\eta_0$ (blue line) and $\eta_0$(green line)(b) for different values of membrane's reflectivity: $r_c=0.98$(red line), $r_c=0.99$ (blue line) and $r_c=0.9999$(green line).\\
\\
\textbf{Fig.6:} The effective temperature of Eq.(\eqn{temp}) versus
the normalized effective detuning for different values of  LDP: $0.85\eta_0$(red line), $0.9\eta_0$ (blue line) and $\eta_0$(green line).\\
\\
\\
\\
\\
\vskip 36cm

\begin{figure}
    \begin{center}
   \includegraphics[width=4in]{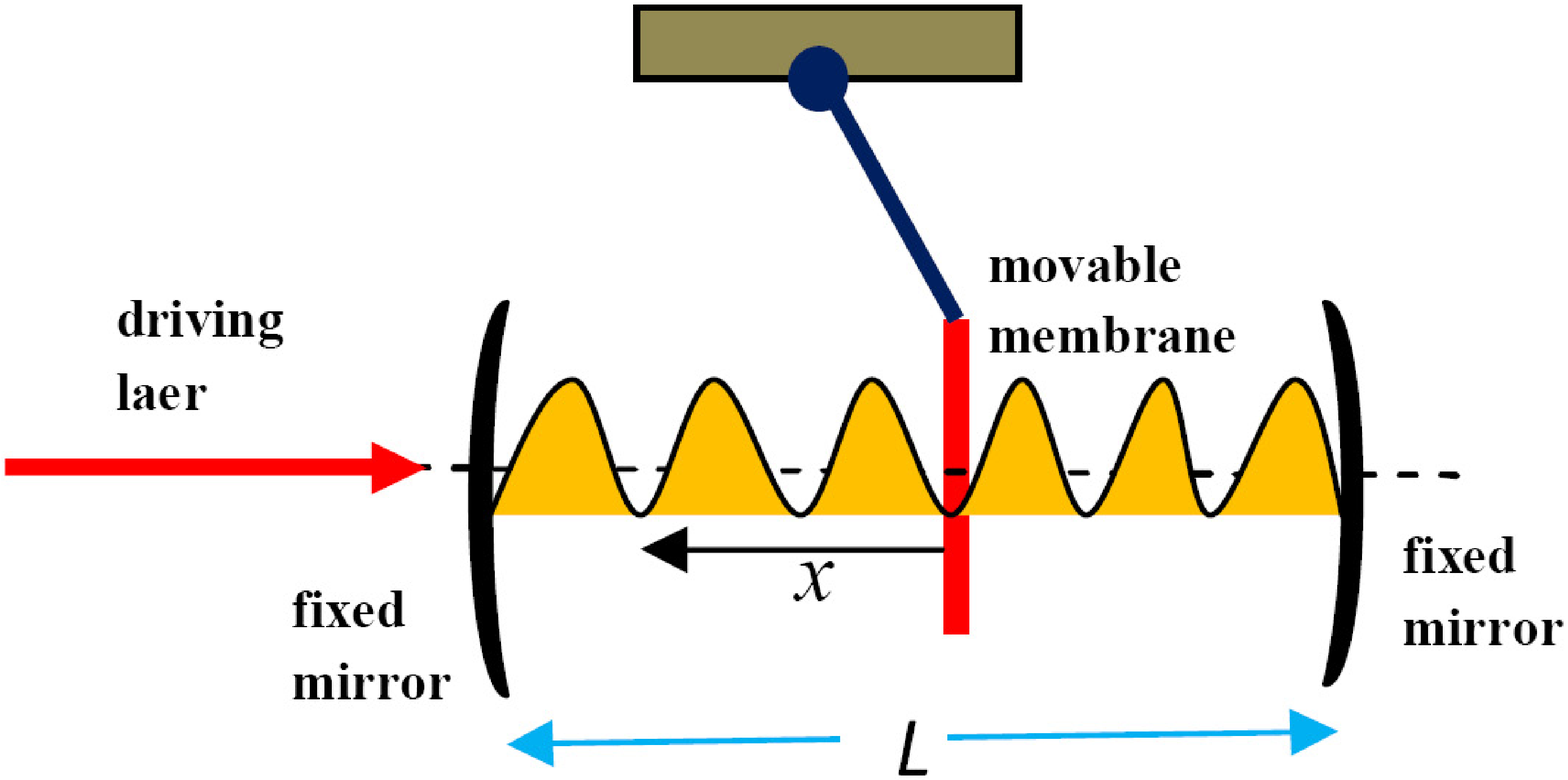}
    \caption{}
    \end{center}
\end{figure}
\begin{figure}
    \begin{center}
    \includegraphics[width=4in]{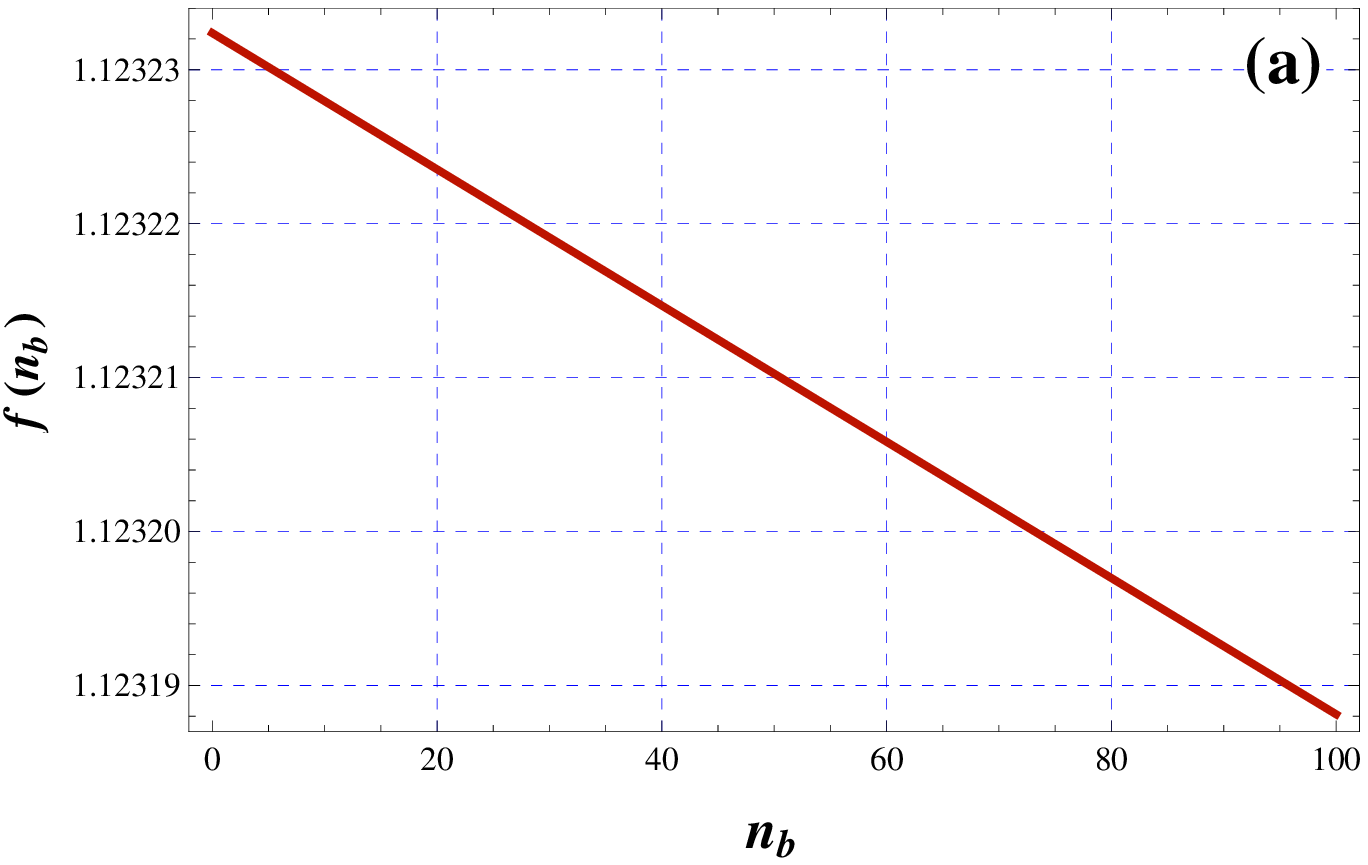}
     \includegraphics[width=4in]{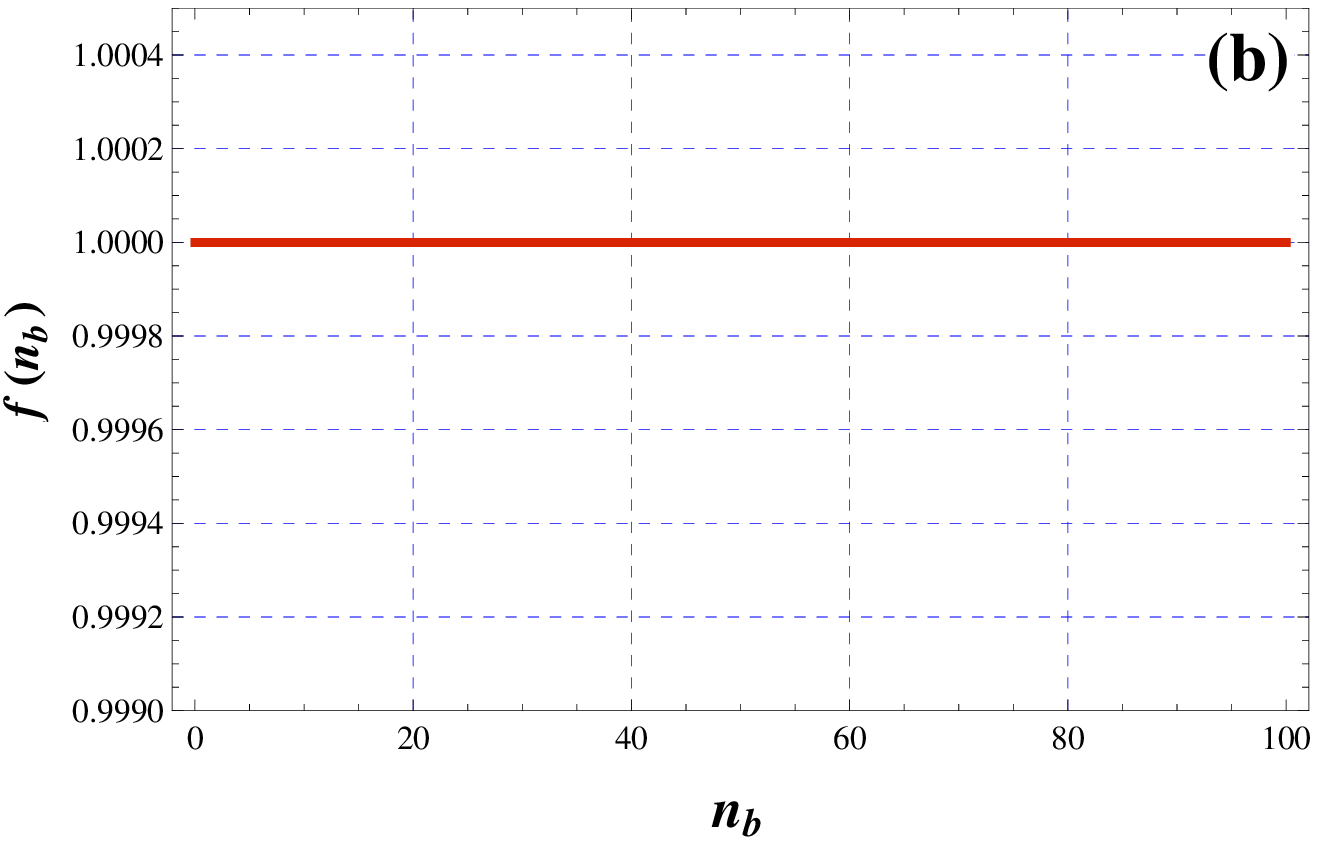}
      \caption{}
       \end{center}
\end{figure}
\begin{figure}
    \begin{center}
    \includegraphics[width=4in]{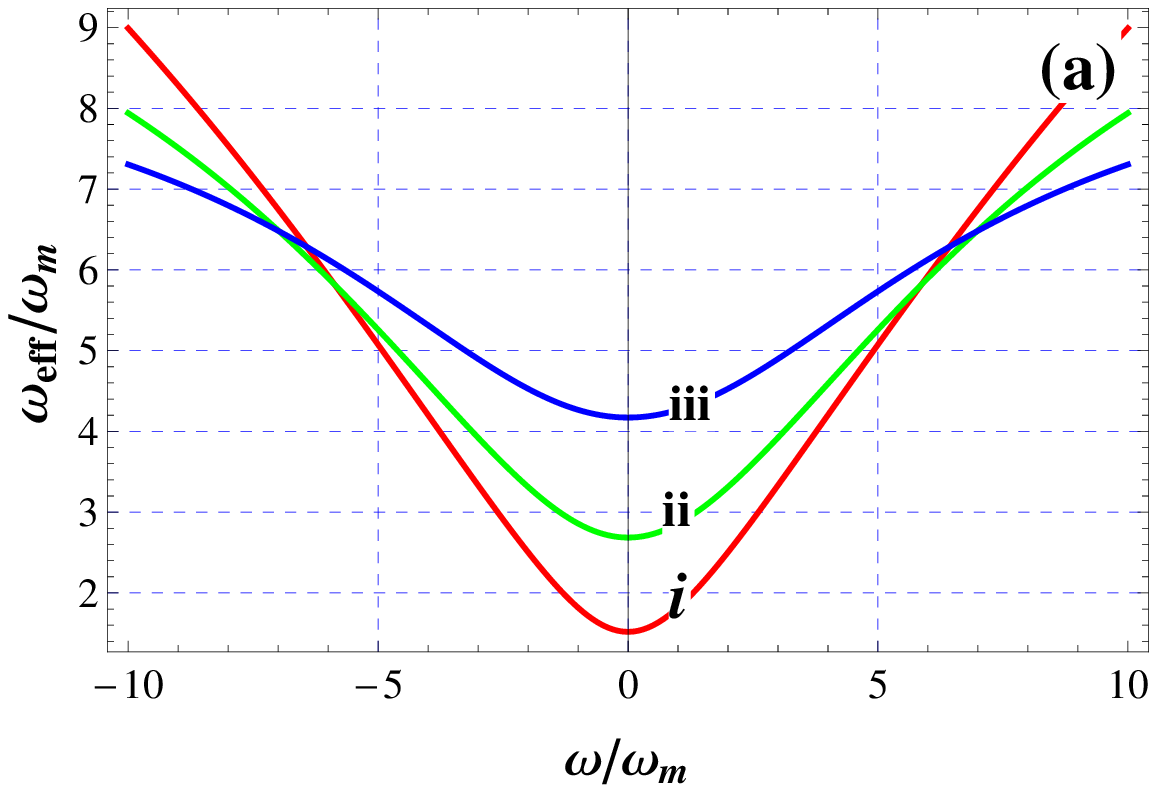}
 \includegraphics[width=4in]{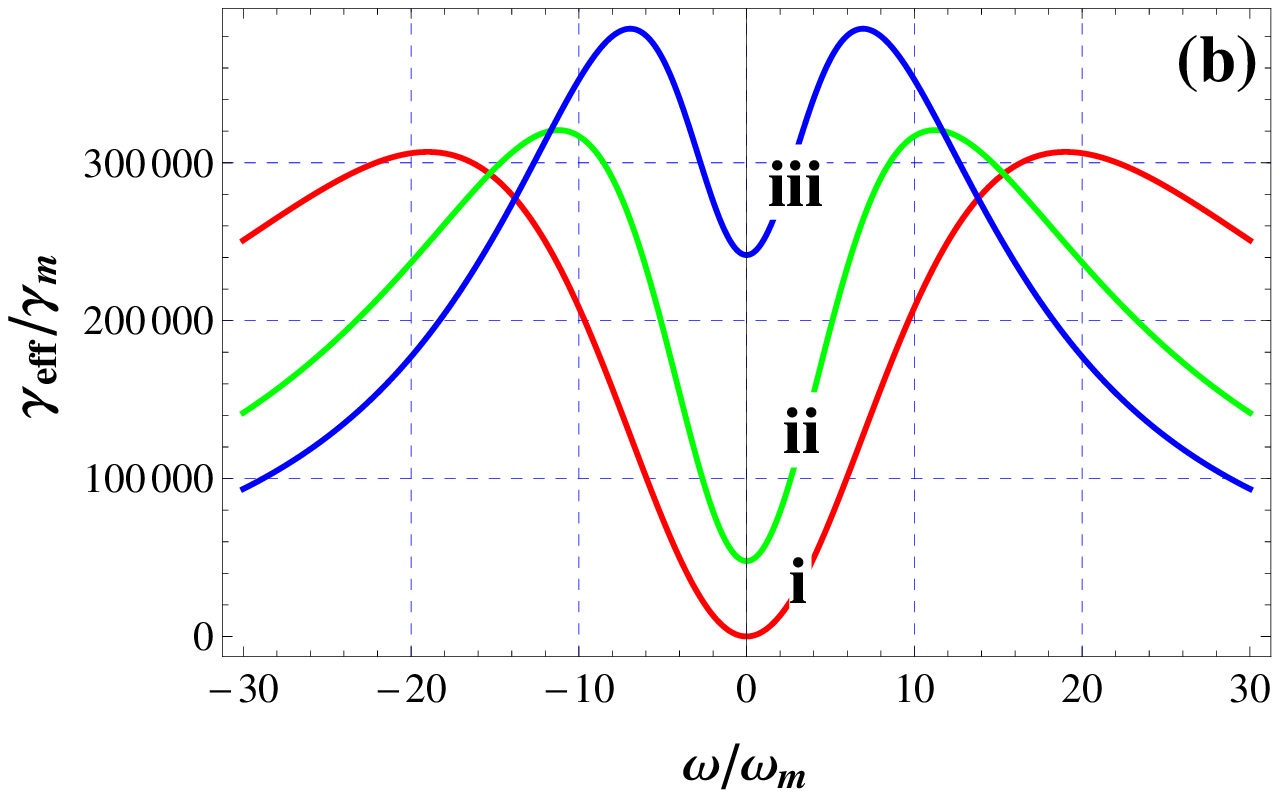}
  \caption{}
   \end{center}
\end{figure}
\begin{figure}
    \begin{center}
    \includegraphics[width=4in]{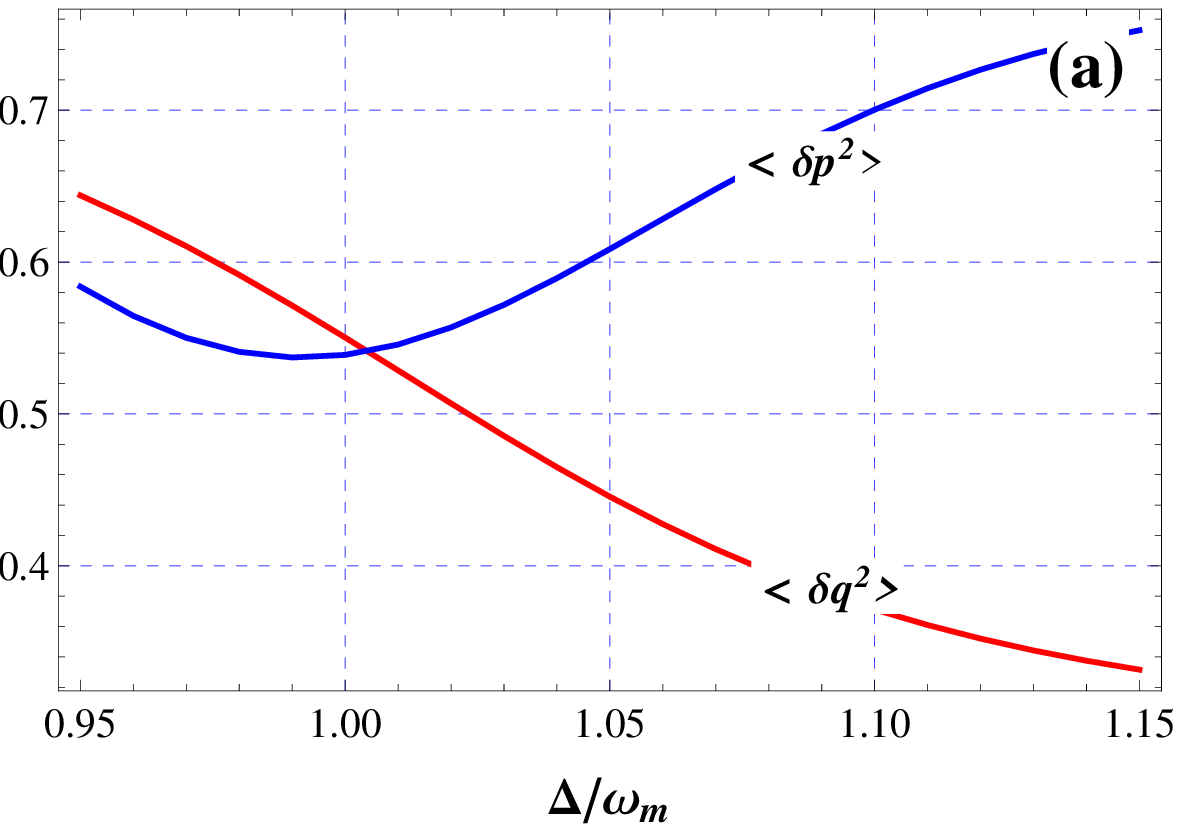}
 \includegraphics[width=4in]{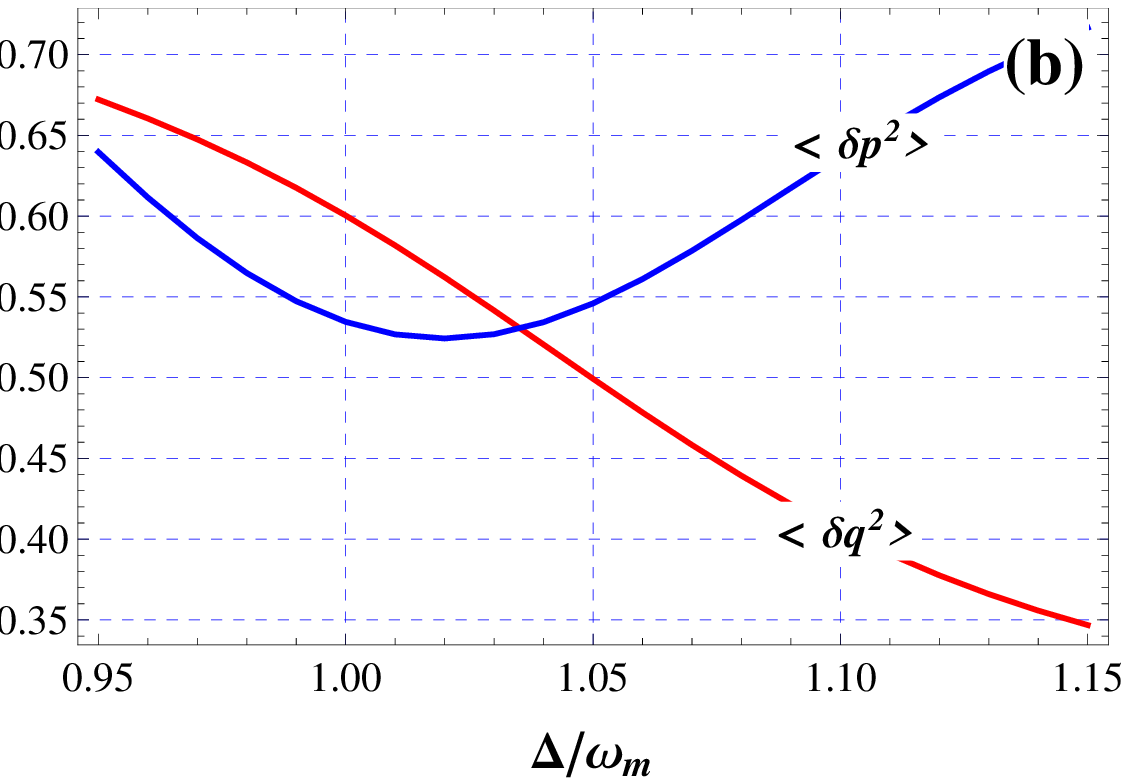}
 \includegraphics[width=4in]{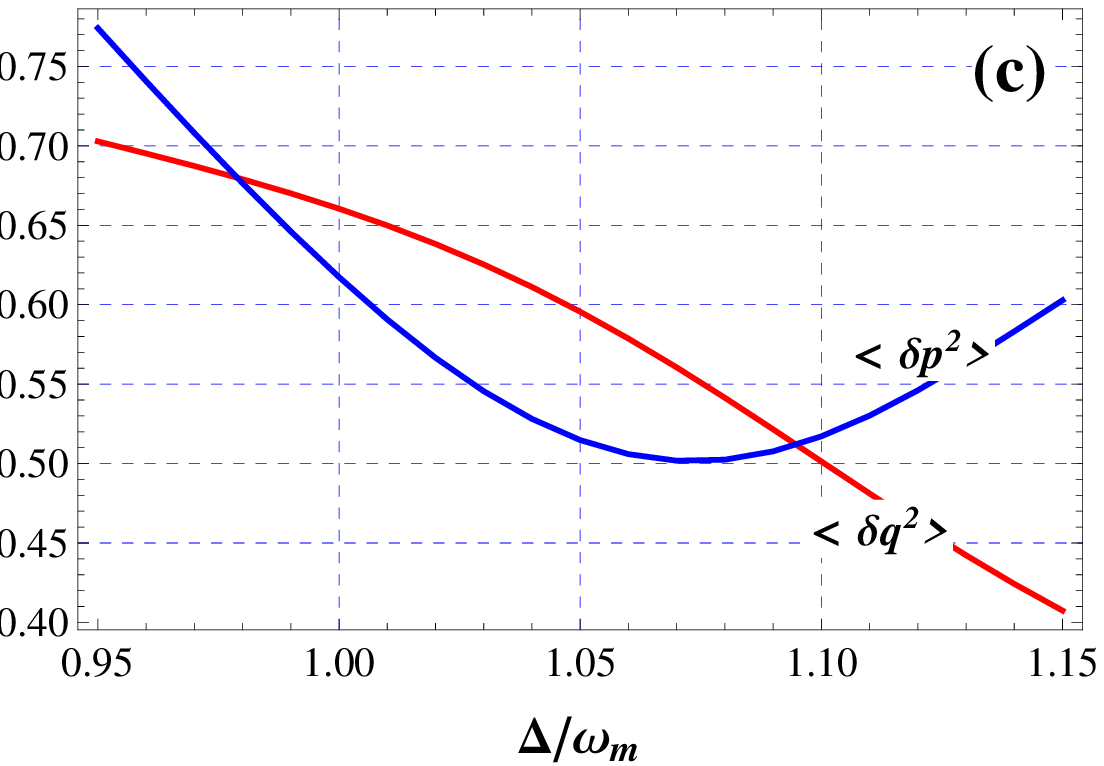}
  \caption{}
   \end{center}
\end{figure}
\begin{figure}
    \begin{center}
    \includegraphics[width=4in]{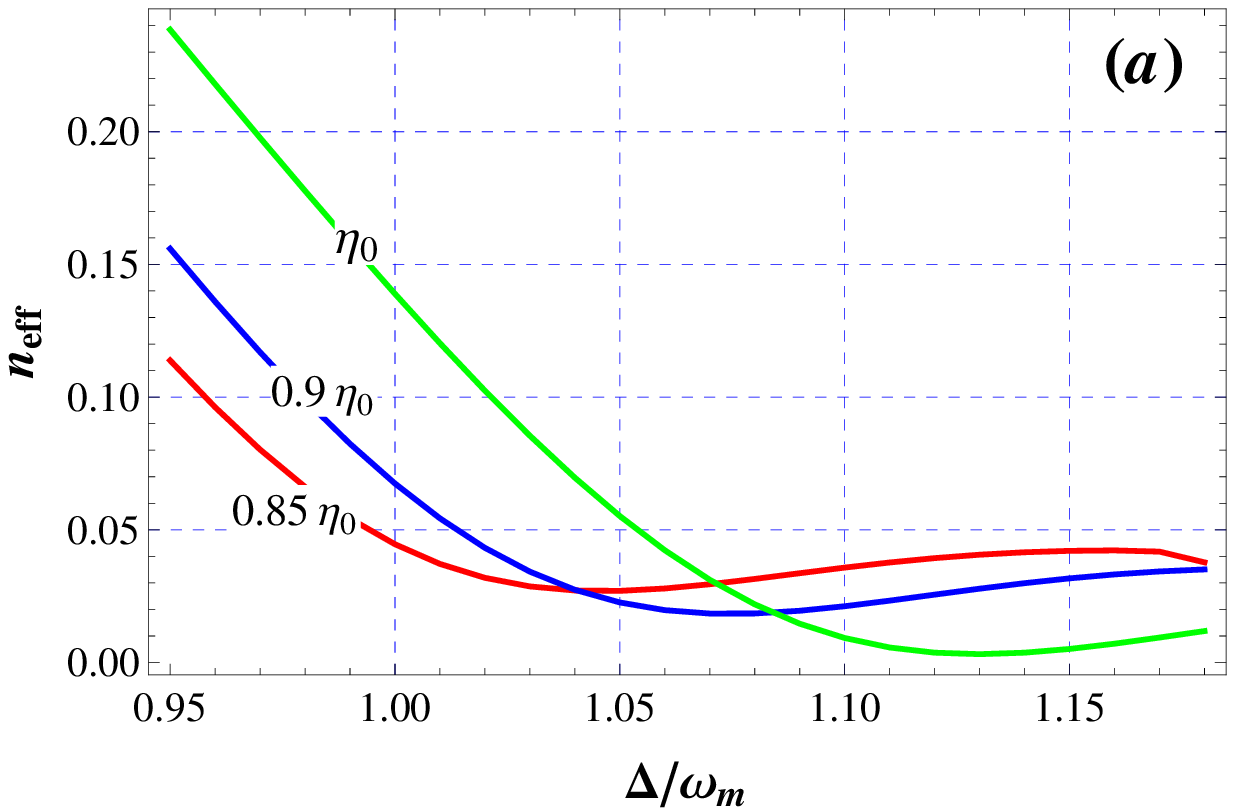}
    \includegraphics[width=4in]{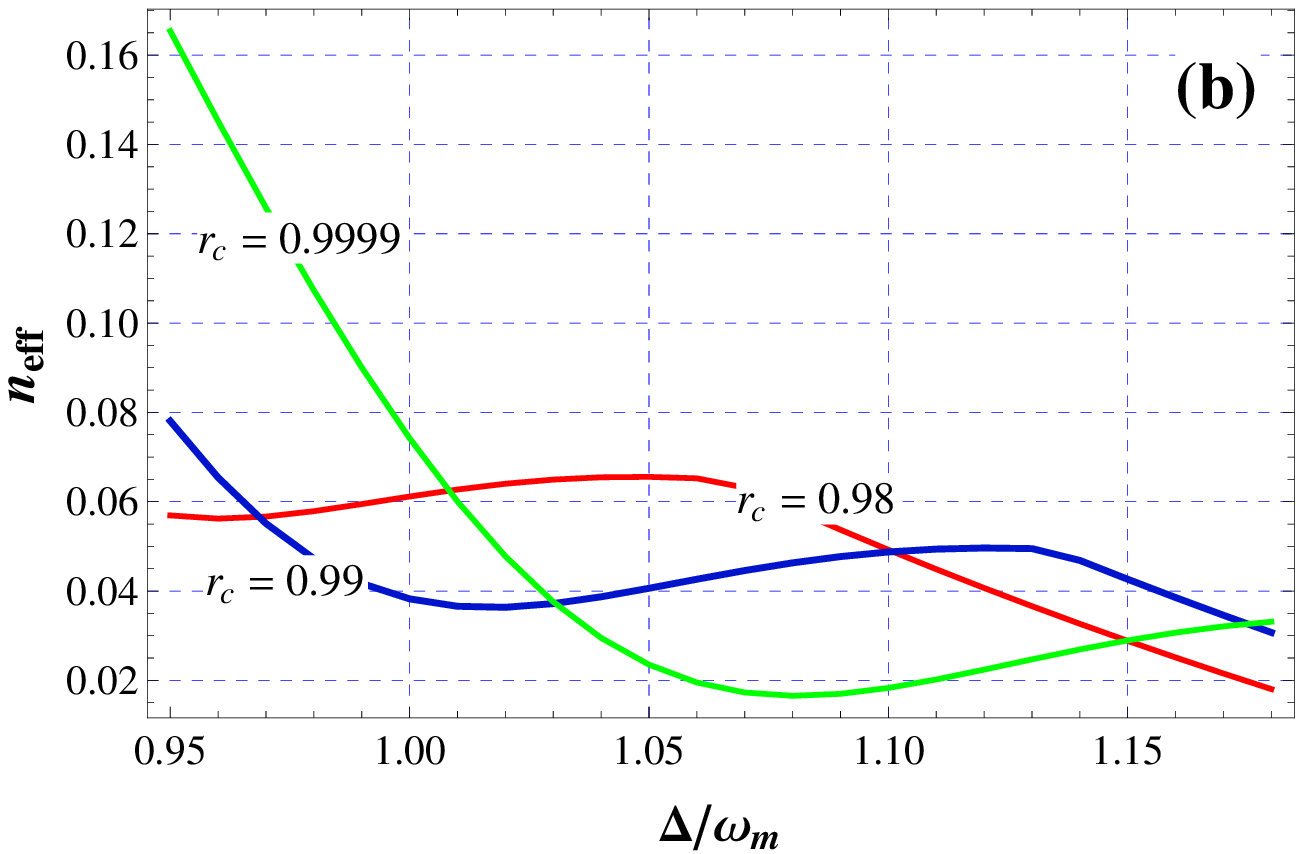}
  \caption{}
   \end{center}
\end{figure}
\begin{figure}
    \begin{center}
    \includegraphics[width=4in]{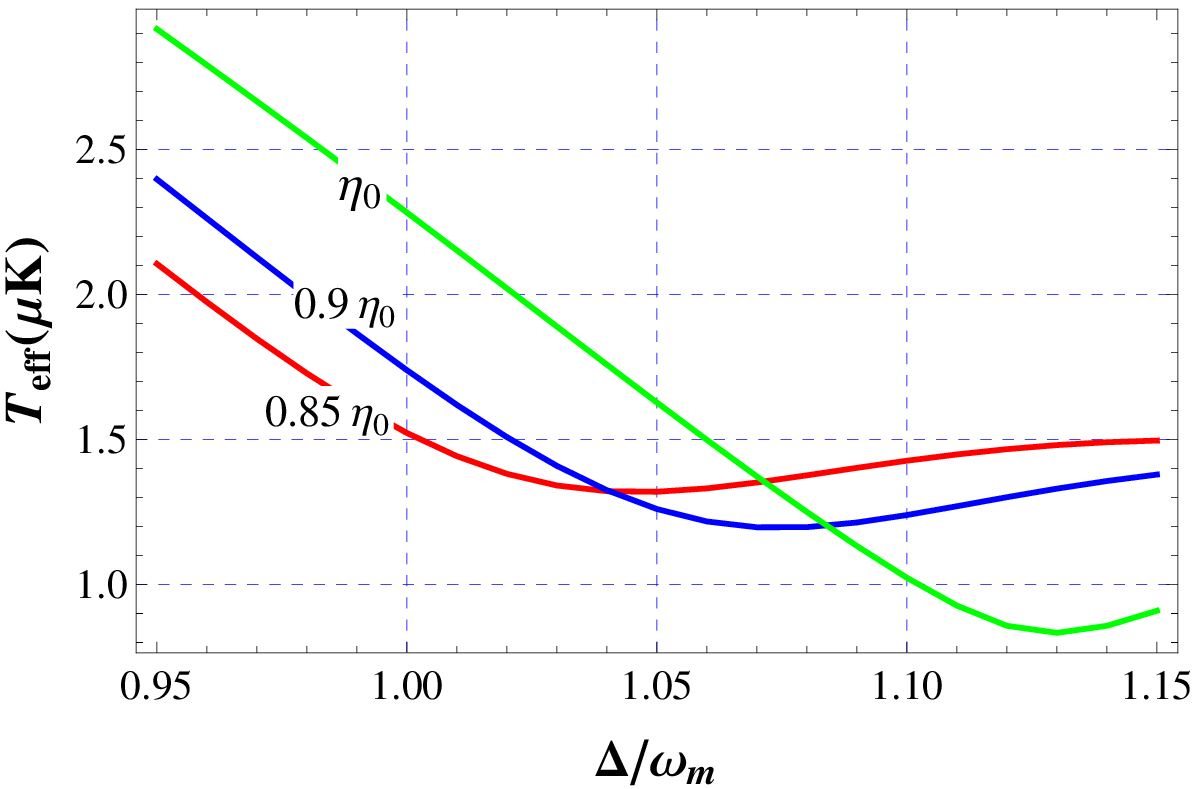}
  \caption{}
   \end{center}
\end{figure}
\end{document}